\documentclass[fleqn,twoside]{article}
\usepackage{espcrc2}
\usepackage{epsfig}
\usepackage[figuresright]{rotating}

\newcommand{\AmS}{{\protect\the\textfont2
  A\kern-.1667em\lower.5ex\hbox{M}\kern-.125emS}}

\title{Non-equilibrium properties of the $S=\frac{1}{2}$ Heisenberg model
       in a time-dependent magnetic field}

\author{V. Turkowski$^{a}$
                     \thanks{Corresponding author.Tel.: +351-21-8419107; 
                             fax: +351-21-8419143.
                             {\it E-mail address:} vturk@cfif.ist.utl.pt},
         V.R. Vieira$^{a}$
     and P.D. Sacramento\address{CFIF, Instituto Superior Tecnico,
                                 Av.Rovisco Pais, 1049-001 Lisbon, Portugal}
       }
       
\begin{document}

\begin{abstract}
The time-dependent behavior of the Heisenberg model 
in contact with a phonon heat bath and in 
an external time-dependent magnetic field is studied by means
of a path integral approach. The action of the phonon heat bath is taken into
account up to the second order in 
%perturbation theory.
the coupling to the heath bath.
It is shown that there is a minimal value of the magnetic field
below which the average magnetization of the system
does not relax to equilibrium when the external magnetic field is flipped.
This result is in qualitative agreement with the
mean field results obtained within $\phi^{4}$-theory.

\vspace{1pc}


\end{abstract}

\maketitle

\section{INTRODUCTION}

The problem of the behavior of a real quantum system is complicated
in many cases by the fact that it is coupled to the environment
degrees of freedom. Therefore, the adequate description of the 
non-equilibrium time-dependent properties of quantum systems in contact
with a thermal reservoir and in external time-dependent fields is 
an interesting and important problem of modern theoretical physics.
In particular, this description is necessary for studying stochastic
processes in mathematical physics \cite{Feynman1,Caldeira1,Schuster1,Zanella1}
and for studying the effective potentials at finite temperatures 
in particle physics \cite{Dolan1}. In condensed matter physics examples
of this type of problems are two-state systems coupled to a dissipative 
environment \cite{Leggett1}, spin systems in a heat bath environment
\cite{Caldeira1,Vieira1,Vieira2,Villain1,Politi1,Garanin1,Prokofev1,Prokofev2},
superconductors in contact with a heat bath
\cite{Abrahams1,Stoof1,Aitchison1,Aitchison2,Sharapov1,Sharapov2,Alamoudi1}
and many others.

The standard method for studying quantum systems out of equilibrium
is the closed-time-path Green function formalism \cite{Schwinger1,Keldysh1}.
For spin systems in general, this method can't be applied directly,
since the commutators of arbitrary spin operators ${\vec S}$ 
are not $c$-numbers, and for this case a standard Wick theorem does not
exist \cite{Doniach1}. However, for the particular case of $S=\frac{1}{2}$,
when the operators anti-commute, the corresponding path integral technique
has been developed \cite{Vieira1,Berezin1,Vieira3}. This technique
is based essentially on standard field theory methods with
the introduction of Majorana fermions.
In this paper we study the non-equilibrium properties of the $S=1/2$
ferromagnetic
Heisenberg model in contact with a phonon heat bath by means of this method
(the Heisenberg model in equilibrium was studied with this technique
in \cite{Vieira2,Sacramento1}, for example). 

There are several reasons
for studying spin systems out of equilibrium. For example, recently discovered
big magnetic molecules \cite{Kahn1,Gatteschi1,Caneschi1,Sessoli1} 
show relaxation times which are extremely large (see also \cite{Paulsen1},
where the experimental results for the case of external magnetic field
are described). Theoretical studies of this phenomenon were
performed in \cite{Villain1,Politi1,Prokofev1}, but more quantitative
analysis is still lacking. 

The behavior of a spin system in the presence of an external
magnetic field flip ${\vec H} \rightarrow -{\vec H}$ is also
an interesting problem. In particular, the question whether there
exists a critical value of the magnetic field below which
the system does not relax to equilibrium and is in a quasi-periodic regime,
as it takes place in the case of a $\phi^{4}$-theory 
\cite{Cao1,Cao2}, is interesting.

In this paper the questions mentioned above are studied in the case
of the Heisenberg model coupled to a phonon heat bath
and in an external time-dependent magnetic field. We consider
the interacting spin system in a magnetic field on the mean field level,
and the phonon heat bath is studied as a perturbation up to the
second order in the spin-phonon coupling.
We introduce the formalism specific to the spin problem and apply it
to several examples of a single spin in a magnetic field and then take
the leading approximation to study interacting spins.

\section{FORMALISM}

The Heisenberg Hamiltonian with a spin-phonon coupling
in an external magnetic field can be written as:
\begin{eqnarray}
{\hat H}=-\frac{1}{2}\sum_{ij}J_{ij}{\vec S}_{i}{\vec S}_{j}-
   \sum_{i}{\vec H}_{i}{\vec S}_{i} \nonumber \\
+\sum_{qi}c_{qi}{\vec S}_{i}{\vec X}_{qi}+
 \frac{1}{2}\sum_{qi}
(\frac{{\vec p}_{qi}^{2}}{m_{qi}}+m_{q}\omega_{qi}^{2}
{\vec X}_{qi}^{2})
\label{H}
\end{eqnarray}
where $J_{ij}>0$ is the nearest neighbor ferromagnetic coupling,
$i,j$ are site coordinates, $q$ is a bath mode with coordinate
${\vec X}_{qi}$, momentum ${\vec p}_{qi}$ and frequency
$\omega_{qi}$ and $c_{qi}$ is the spin-phonon coupling.

The normalized generating functional of the system is
$$
{\cal Z}=\frac{Tr\left[{\hat T}
         e^{-\int_{0}^{\beta_{f}}{\hat H}_{b}d\tau}
         e^{-\int_{0}^{\beta_{i}}{\hat H}_{s}d\tau}
         e^{-i\int_{t_{f}}^{t_{i}}{\hat H}dt}
         e^{-i\int_{t_{i}}^{t_{f}}{\hat H}dt}
\right]
}
{Tr\left[{\hat T}e^{-\int_{0}^{\beta_{f}}{\hat H_{b}}d\tau}\right]
 Tr\left[{\hat T}e^{-\int_{0}^{\beta_{i}}{\hat H_{s}}d\tau}\right]
},
$$
\begin{eqnarray}
\label{Z1}
\end{eqnarray}
where ${\hat H}_{s}$ and ${\hat H}_{b}$ are the spin and bath 
parts of the Hamiltonian, respectively.
Initially the spins and the phonons are decoupled. The spin system is at
a temperature $T_{i}=\frac{1}{\beta_{i}}$ and the phonon bath
is at a temperature $T_{f}=\frac{1}{\beta_{f}}$. At time $t_{i}$
the spin-phonon coupling is turned on and it is assumed that
the spin system will reach thermodynamic equilibrium with the phonons.
In the closed-time path Green function formalism the numerator 
of the generating functional can be written in this case in terms of the path
integral:                                
\begin{eqnarray}
\int D{\vec \zeta}  D{\vec X}
\exp\left(-i\int_{C}dt
\left[-\frac{i}{2}\sum_i{\vec \zeta}_{i}\frac{d}{dt}{\vec \zeta}_{i}
\right.\right.
\nonumber \\ 
\left.\left.
-\frac{1}{2}\sum_{ij}J_{ij}{\vec S}_{i}{\vec S}_{j}
-\sum_i{\vec S}_{i}{\vec H}_{i}
\right.\right.
\nonumber \\ 
\left.\left.
+\sum_{qi}c_{qi}{\vec S}_{i}{\vec X}_{qi}+
\sum_{qi}{\vec X}_{qi}{\tilde D}_{qij}^{-1}{\vec X}_{q,j}\right]
\right)
\label{Z2}
\end{eqnarray}
where ${\tilde D}$ is the phonon propagator and the time integration
contour $C$ is presented in Fig.1.

\begin{figure}% fig 1
\begin{center}
\epsfig{file=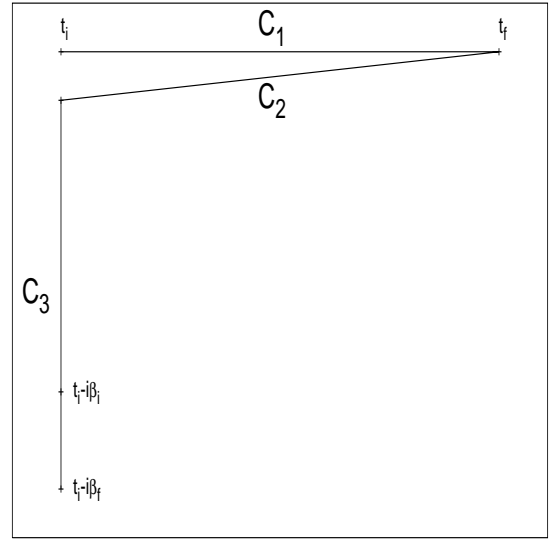,height=3.0in,width=3.0in,angle=270}
\end{center}
\caption{Integration contour for the time variable. The direction
is $t_{i}\rightarrow t_{f}\rightarrow t_{i}
\rightarrow t_{i}-i\beta_{i},t_{i}-i\beta_{f}$.}
\label{foobar:fig1}
\end{figure}

In (\ref{Z2}) a Majorana fermion operator ${\vec \zeta}_{i}$ is introduced.
In the case of the spin $\frac{1}{2}$ the following relation holds
\cite{Vieira3,Vieira1}:
\begin{equation}
{\vec S}_{i}=-\frac{i}{2}{\vec \zeta}_{i}\times {\vec \zeta}_{i}.
\label{Majorana}
\end{equation}
After the integration over the phonon modes, the numerator of the generating functional 
(\ref{Z2}) becomes
\begin{eqnarray}
%{\cal Z}=
%\frac{
\int D{\vec \zeta}
\exp\left(-i\int_{C}dt
\left[-\frac{i}{2}\sum_i{\vec \zeta}_{i}\frac{d}{dt}{\vec \zeta}_{i}
\right.\right.
\nonumber \\
\left.\left.
-\frac{1}{2}\sum_{ij}{\vec S}_{i}{\tilde J}_{ij}{\vec S}_{j}
-\sum_i{\vec S}_{i}{\vec H}_{i}
\right]
\right)
\nonumber
%}
%{
%Tr{\hat T}\exp\left( -\int_{0}^{\beta_{i}}{\hat H_{s}}d\tau\right)
%}
\label{Z3}
%\end{equation}
\end{eqnarray}
and the trace of the heath bath in the denominator is cancelled.
The presence of the phonon bath environment 
is given by the Feynman and Vernon influence functional 
\cite{Feynman1,Caldeira1}. Its effect is to change
%results in changing 
the spin-spin coupling to an effective coupling 
\begin{equation}
{\tilde J}_{ij}(t-t')\rightarrow 
J_{ij}\delta (t-t')+i\Delta_{ij}(t-t').
\label{Jtilde}
\end{equation}
The function $\Delta_{ij}(t-t')$ is %\cite{Feynman1,Caldeira1}
$$
\Delta_{ij}(t-t')=\alpha_{ij}(t-t')\theta (t-t') +
                  \alpha_{ij}^{*}(t-t')\theta (t'-t),
$$
$$
\alpha_{ij}(t-t')=\sum_{q}\frac{c_{qi}^{*}c_{qj}}{2m_{q}\omega_{q}}
\left[\frac{e^{i\omega_{q}(t-t')}}{e^{\beta_{f}\omega_{q}}-1}+
 \frac{e^{-i\omega_{q}(t-t')}}{1-e^{-\beta_{f}\omega_{q}}}
\right]
$$
and it contains all the effective properties of the bath environment.
It can be approximated in the standard way \cite{Caldeira1}:
$$
\alpha_{ij}(t-t')\simeq 
\delta_{ij}A\int_{0}^{\infty}d\omega \omega^{s} 
e^{-\omega /\omega_{c}}
$$
$$
\times\left[\frac{e^{i\omega (t-t')}}{e^{\beta_{f}\omega}-1}+
 \frac{e^{-i\omega (t-t')}}{1-e^{-\beta_{f}\omega}}
\right],
$$
where $A,s$ and $\omega_{c}$ are effective 
phonon density of states parameters .
In particular, $A$ is an effective spin-phonon coupling,
$\omega_{c}$ is the phonon energy cut-off, and
$s$ describes the ``subohmic'', ``ohmic'' and
``superohmic'' cases, when \mbox{$0<s<1$}, $s=1$ and $s>1$, respectively. 
Below we shall consider separately these different cases. 

To decouple the interaction between the spins let us make the
Hubbard-Stratonovich transformation:
\begin{eqnarray}
{\cal Z}=\int\frac{D{\vec \zeta}D{\vec \phi}}{\sqrt{det(2\pi{\tilde J})}}
\exp\left(-i\int_{C}dt\left[
-\frac{i}{2}\sum_i{\vec \zeta}_{i}\frac{d}{dt}{\vec \zeta}_{i}\right.\right.
\nonumber \\
\left.\left.
+\frac{1}{2}\sum_{ij}{\vec \phi}_{i}{\tilde J}_{ij}^{-1}{\vec \phi}_{j}-
\sum_i{\vec \phi}_{i}{\vec S}_{i}-\sum_i{\vec H}_{i}{\vec S}_{i}\right]
\right), 
\label{HS1}
\end{eqnarray}
and with the shift 
${\vec \phi}_{i}\rightarrow {\vec \phi}_{i}-{\vec H}_{i}$:
\begin{eqnarray}
{\cal Z}=\int\frac{D{\vec \zeta}D{\vec \phi}}{\sqrt{det(2\pi{\tilde J})}}
\exp\left(-i\int_{C}dt\left[
-\frac{i}{2}\sum_i{\vec \zeta}_{i}\frac{d}{dt}{\vec \zeta}_{i}\right.\right.
\nonumber \\
\left.\left.
+\frac{1}{2}\sum_{ij}({\vec \phi}_{i}-{\vec H}_{i}){\tilde J}_{ij}^{-1}
            ({\vec \phi}_{j}-{\vec H}_{i})-
\sum_i{\vec \phi}_{i}{\vec S}_{i}\right]
\right), 
\label{HS2}
\end{eqnarray}

In the case of the substitution (\ref{Majorana})
the effective fermion Hamiltonian is
$$
{\hat H}_{eff}=-\sum_{i}{\vec \phi}_{i}{\vec S}_{i}=
\frac{i}{2}\sum_{ilmn}\phi_{i}^{l}\varepsilon^{lmn}
\zeta_{i}^{m}\zeta_{i}^{n},
$$
where $l,m,n$ are vector components. The action in the exponent
of (\ref{HS2}) is quadratic on the $\zeta$-variables.
Therefore it can be formally integrated,
\begin{eqnarray}
{\cal Z}=\int\frac{D{\vec \phi}}{\sqrt{det(2\pi{\tilde J})}}
\nonumber \\
\times e^{-\frac{i}{2}\int_{C}dt
\sum_{ij}({\vec \phi}_{i}-{\vec H}_{i})
{\tilde J}_{ij}^{-1}({\vec \phi}_{j}-{\vec H}_{j})}
\prod_iK({\vec \phi_i}),
\label{Z5}
\end{eqnarray}
where
\begin{equation}
K({\vec \phi_i})=\int D{\vec \zeta}
\exp\left(-\int_{C}dt\left[
\frac{1}{2}{\vec \zeta}_{i}\frac{d}{dt}{\vec \zeta}_{i}+
i{\hat H}_{eff}\right]\right)
\label{K}
\end{equation}
is a function of the order parameter ${\vec \phi}$. 
The functional $K({\vec \phi_i})$ is the Helmholtz free energy
for a single spin in an external effective field %${\vec H}_{eff}=$
${\vec \phi_i}$.

If the external field is uniform in space and time, the propagator of the  
$\zeta$-fields 
\begin{equation}
G_{ij}^{ab}(t,t')=-i<{\hat T}\zeta_{i}^{a}(t)\zeta_{j}^{b}(t')>
\label{GF}
\end{equation}
can be obtained analytically. %(most conventionally in the spherical basis). 
%The expression for the Green's function $G (t_{1},t_{2})$ 
In the case of an external
magnetic field ${\vec H}=(0,0,H^{z})$, 
%can be obtained analytically 
this can be done most conventionally in the spherical basis with unit vectors 
${\vec a}_{\pm}=\frac{1}{\sqrt{2}}({\vec e}_{x}\pm i{\vec e}_{y})$, 
${\vec a}_{0}={\vec e}_{z}$ \cite{Vieira2}.
%This function is 
It becomes diagonal 
%in this case 
and satisfies the equation
$$
(i\frac{d}{dt}-\alpha H^{z})G_{\alpha}(t-t')=\delta (t-t'),
$$
where $\alpha =\pm ,0$ denotes the spherical components.
%according to the definition of the axes vectors.
The times $t$ and $t'$ take values on the contour $C$, and therefore
$G$ can be presented 
%for different times on the contour 
as
\begin{eqnarray}
G_{\alpha}(t-t')=G_{\alpha}^{>}(t-t')\theta_{C}(t-t')
\nonumber \\
+G_{\alpha}^{<}(t-t')\theta_{C}(t'-t),
\label{G}
\end{eqnarray}

\begin{figure}[h!]% fig 2
\begin{center}
\epsfig{file=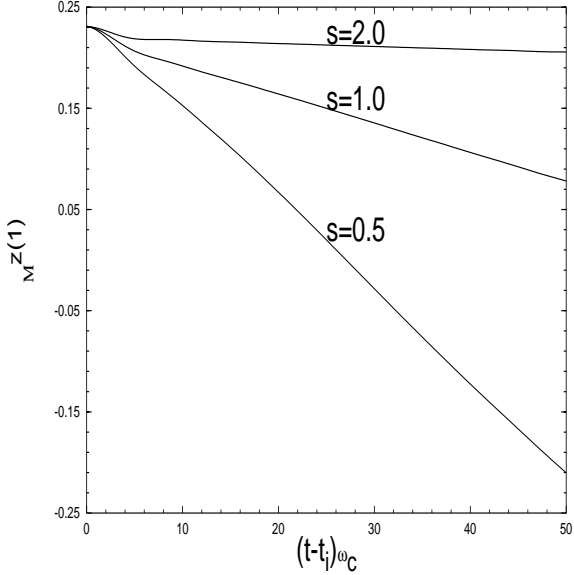,height=3.0in,width=3.0in,angle=270}
\end{center}
\caption{Time-dependence of one-loop magnetization for one spin after 
magnetic field flipping at different values
of $s$ and with 
$A=0.025, H/\omega_c=0.1, T_i/\omega_c=0.1, T_f/\omega_c=0.02$.}
\label{foobar:fig2}
\end{figure}
where $\theta_{C}(t'-t)$ is the theta-function on the contour $C$.

The solution of (\ref{GF}) in the case of a time-independent magnetic field
${\vec H}$ with the Kubo-Martin-Schwinger boundary
conditions \cite{Kubo} 
$$
G_{\alpha}^{>}(t-i\beta_{i}-t')=-G_{\alpha}^{<}(t-t')
$$
is
\begin{equation}
G_{\alpha}^{>}(t-t')=-i\frac{1}{e^{-\beta_i\alpha H^{z}}+1}
e^{-i\alpha H^{z}(t-t')},
\label{G>}
\end{equation}
\begin{equation}
G_{\alpha}^{<}(t-t')=i\frac{1}{1+e^{\beta_i\alpha H^{z}}}
e^{-i\alpha H^{z}(t-t')}.
\label{G<}
\end{equation}

However,
in general, since the field may be both time and space dependent there 
is no simple analytical solution. Therefore, in practice,
the $\zeta$-integration in (\ref{K}) can be performed
using the discrete path integral approximation, with the time contour 
divided on a finite number $N$ of time intervals \cite{Vieira1}
and Eq.(\ref{K}) written as:
\begin{eqnarray}
K(\phi )=\frac{1}{2^{3N/2}}\int ... \int 
e^{\zeta^{T}{\hat A}\zeta-
i\sum_{k=0}^{N-1}{\hat H}_{eff}(t_{k})\Delta t}
\nonumber \\
\times d\zeta_{1}...d\zeta_{N},
\label{Kdiscrete1}
\end{eqnarray}
where
$$
{\hat A}=\left(
\begin{array}{ccccc}
\ \ \ \ 0\  \ -{\hat 1} \  \ \ \ {\hat 1} \ \ ...\ \ -{\hat 1} \\
\ \ \  {\hat 1}\ \ \ 0 \  \ -{\hat 1} \ \ ...\ \  {\hat 1} \\
-{\hat 1}\  \ \ \ {\hat 1} \  \ \ \  0 \ \ \ \ ...\ \ -{\hat 1} \\
...\  \ \ \ ... \  \ ... \ \ ...\ \ ... \\
\ \ {\hat 1} \  \  -{\hat 1} \  \ {\hat 1} \ \ ... \ \ 0 \\
\end{array}\right),
\ \ \ \ \ \ \ 
\zeta=\left(
\begin{array}{c}
{\vec \zeta}_{1}\\
{\vec \zeta}_{2}\\
...\\
{\vec \zeta}_{N}\\
\end{array}\right),
$$
and ${\hat 0}$ and ${\hat 1}$ are diagonal $3\times 3$ zero-
and unit-matrices, respectively. 
%In short notation the expression (\ref{Kdiscrete1}) can be written as
%\begin{equation}
%K(\phi )=\frac{1}{2^{3N/2}}\int ... \int 
%e^{\frac{i}{2}\zeta^{T}G^{-1}\zeta} 
%d\zeta_{1}...d\zeta_{N},
%\label{Kdiscrete2}
%\end{equation}
The discrete inverse propagator in the presence of a general magnetic field 
${\vec\phi}$ is then 
\begin{equation}
G^{-1}=-2i\left(
\begin{array}{ccccc}
{\tilde C}\  \ -{\hat 1} \  \ {\hat 1} \ \ ...\ \ -{\hat 1} \\
{\hat 1}\  \ \ {\tilde C} \ -{\hat 1} \ \ ...\ \  {\hat 1} \\
-{\hat 1}\  \ {\hat 1} \  \ {\tilde C} \ \ ...\ \ -{\hat 1} \\
...\  \ ... \  \ ... \ \ ...\ \ ... \\
{\hat 1} \  \  -{\hat 1} \  \ {\hat 1} \ \ ... \ \ {\tilde C} \\
\end{array}\right) 
\label{GFdisc}
\end{equation}
with
\begin{equation}
{\hat C}=\left(
\begin{array}{ccc}
\ \ 0\  \ \ \ \ \phi_{i}^{z} \ \ -\phi_{i}^{y} \\
-\phi_{i}^{z} \ \ 0 \ \ \ \ \ \phi_{i}^{x}  \\
\phi_{i}^{y}\  \ -\phi_{i}^{x} \  \ \ \ 0 
\end{array}\right) \times\frac{\Delta t}{2}.
\label{C}
\end{equation}

The integration over the $\zeta$-fields in 
%(\ref{Kdiscrete2})
(\ref{K}) gives
\begin{eqnarray}
{\cal Z}=\frac{1}{2^{3N/2}}
\int\frac{D{\vec \phi}}{\sqrt{det(2\pi{\tilde J})}}
\nonumber \\
\times\exp\left(
-\frac{i}{2}\int_{C}dt
\sum_{ij}({\vec \phi}_{i}-{\vec H}_{i}){\tilde J}_{ij}^{-1}
({\vec \phi}_{j}-{\vec H}_{j})
\right.
\nonumber \\
\left.
+
\frac{1}{2}\sum_iTr\ln iG^{-1}({\vec \phi_i})
\right),
\label{Z6}
\end{eqnarray}
where $G({\vec \phi_i})$ is calculated numerically.
The loop expansion on the field
${\vec \phi}$ in the thermodynamic potential can be performed
to take into account the fluctuations of the order parameter
\cite{Sacramento1}.

Minimization of the thermodynamic potential with respect
to ${\vec \phi}$ gives the saddle-point equation for the order parameter:
\begin{equation}
{\vec \phi}_{i}^{(0)}(t)={\vec H}_{i}(t)+
\int_{C}dt'{\tilde J}_{ij}(t,t'){\vec M}_{j}^{(0)}(t'),
\label{phi}
\end{equation}
\begin{figure}[h!]% fig 3
\begin{center}
\epsfig{file=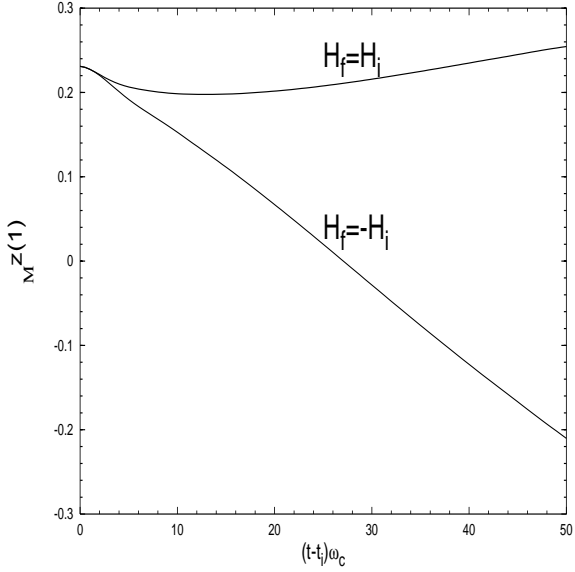,height=3.0in,width=3.0in,angle=270}
\end{center}
\caption{Time-dependence of one-spin magnetization after lowering temperature 
for the cases of magnetic field
flipping and constant magnetic field at $s=0.5, A=0.025, H/\omega_c=0.1, T_i/\omega_c=0.1$ and $T_f/\omega_c=0.02$.}
\label{foobar:fig3}
\end{figure}
where the mean-field magnetization $M_{j}^{l(0)}(t')$ is
$$
M_{j}^{l(0)}(t')=\frac{\delta}{i\delta \phi_{j}^{l}(t')}
\frac{1}{2}Tr\ln iG^{-1}({\vec \phi_j})
$$
$$
= 
\frac{1}{2}\varepsilon^{lsr}G_{jj}^{sr}(t',t').
$$
Taking the time limit properly, one gets
\begin{equation}
M_{j}^{l(0)}(t)=
\frac{1}{4}\varepsilon^{lsr}(G_{jj}^{sr}(t^{+},t)+
G_{jj}^{sr}(t^{-},t)).
\label{M0}
\end{equation}
The self-consistent system of the equations 
(\ref{phi}), (\ref{M0}) together with (\ref{GFdisc})-(\ref{C})
can be solved numerically, giving the ${\vec \phi}$-dependence
of the Green function (\ref{GF}), which allows to analyze
the magnetic properties of the system. In this paper we consider
the spatially homogeneous magnetic field case, so the Green's function
for the $\zeta$-field can be found analytically in some cases of the
field time dependence.

The magnetization $M_{i}^{k(1)}(t)$ calculated to the
order of one-loop in the boson fluctuations is \cite{Sacramento1}:
\begin{equation}
M_{i}^{k(1)}(t)=
\frac{\delta(-\beta F^{(1)})}{\delta H_{i}^{k}(t)},
\label{M1}
\end{equation}
where
$$
\beta F^{(1)}=I({\vec \phi})+\frac{1}{2}Tr\ln{\tilde J}D^{-1}
$$
is the free energy calculated to the order of one
loop in the effective RPA interaction $D$:
\begin{equation}
D_{sj}^{pm-1}(t,t')=
{\tilde J}_{sj}^{pm-1}\delta (t,t')-
i{\bar \chi}_{sj}^{pm}(t,t'),
\label{D}
\end{equation}
with the free magnetic susceptibility
\begin{eqnarray}
{\bar \chi}_{sj}^{pm}(t,t')
\nonumber \\
=-\frac{1}{2}
\varepsilon^{ll_{1}l_{2}}\varepsilon^{mm_{1}m_{2}}
G_{ij}^{l_{2}m_{1}}(t,t')G_{ij}^{m_{2}l_{1}}(t',t).
\label{chi}
\end{eqnarray}
evaluated at the saddle-point (here and below the summation 
over repeated indices is assumed).

The function $I({\vec \phi})$ is the mean-field action
$$
I({\vec \phi})=-\frac{i}{2}\int_{C}dt
\sum_{ij}({\vec \phi}_{i}-{\vec H}_{i}){\tilde J}_{ij}^{-1}
({\vec \phi}_{j}-{\vec H}_{j})
$$
$$
-\frac{1}{2}\sum_iTr\ln iG^{-1}({\vec \phi_i})
$$
The expression for $M^{(1)}$ is
\begin{eqnarray}
M_{i}^{k(1)}(t)=M_{i}^{k(0)}(t)
\nonumber \\
+[\delta_{is}\delta^{nr}\delta (t_{2}-t_{3})
+i{\bar \chi}_{ib}^{na}(t_2,t_4)D_{bs}^{ar}(t_{4},t_{3})]
\nonumber \\
\times B_{sij}^{rnl}(t_{3},t_{2},t_{1})
D_{ji}^{lk}(t_{1},t),
\label{M1detail}
\end{eqnarray}
where the third order vertex is
\begin{eqnarray}
B_{sij}^{rnl}(t_{3},t_{2},t_{1})
\nonumber \\
=\frac{i}{2}\frac{\delta{\bar \chi}_{si}^{rn}(t_3,t2)}{\delta \phi_{j}^{l}(t_1)}
=\frac{1}{4}\varepsilon^{rr_1r_2}\varepsilon^{nn_1n_2}\varepsilon^{ll_{1}l_{2}}
\nonumber \\
\times
(G_{ji}^{l_{2}n_{1}}(t_{1},t_{2})
 G_{is}^{n_{2}r_{1}}(t_{2},t_{3})
 G_{sj}^{r_{2}l_{1}}(t_{3},t_{2})
\nonumber \\
+ G_{js}^{l_{2}r_{1}}(t_{1},t_{3})
 G_{si}^{r_{2}n_{1}}(t_{3},t_{2})
 G_{ij}^{n_{2}l_{1}}(t_{2},t_{1})).
\label{B}
\end{eqnarray}
Let us note that the functions $G, {\bar \chi}$ and $B$
are diagonal on site indices. The Green's functions $G$ in these expressions
are calculated self-consistently by means of equations (\ref{phi}) and 
(\ref{M0}).

\begin{figure}% fig 4
\begin{center}
\epsfig{file=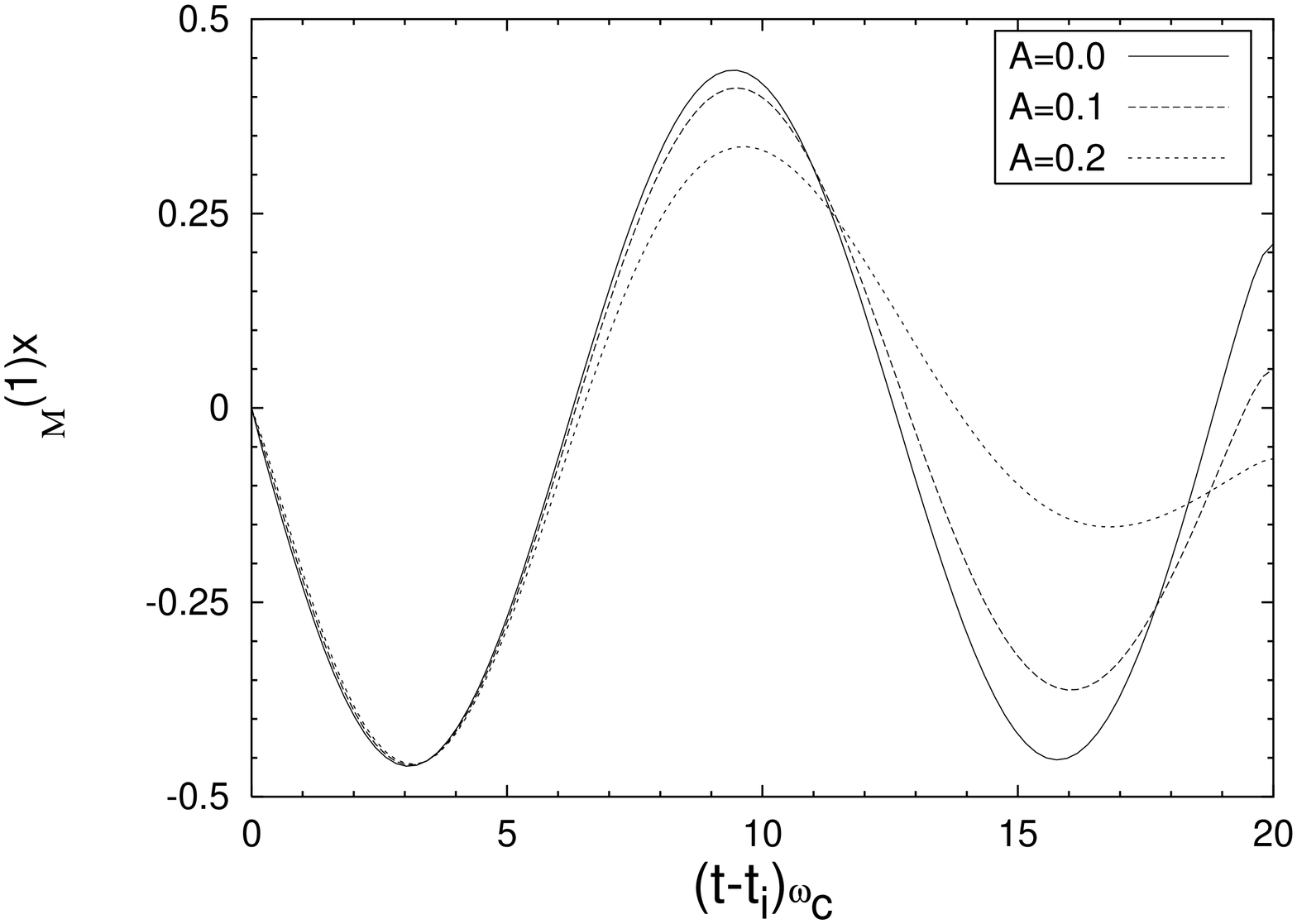,height=2.5in,width=2.5in,angle=270}
\epsfig{file=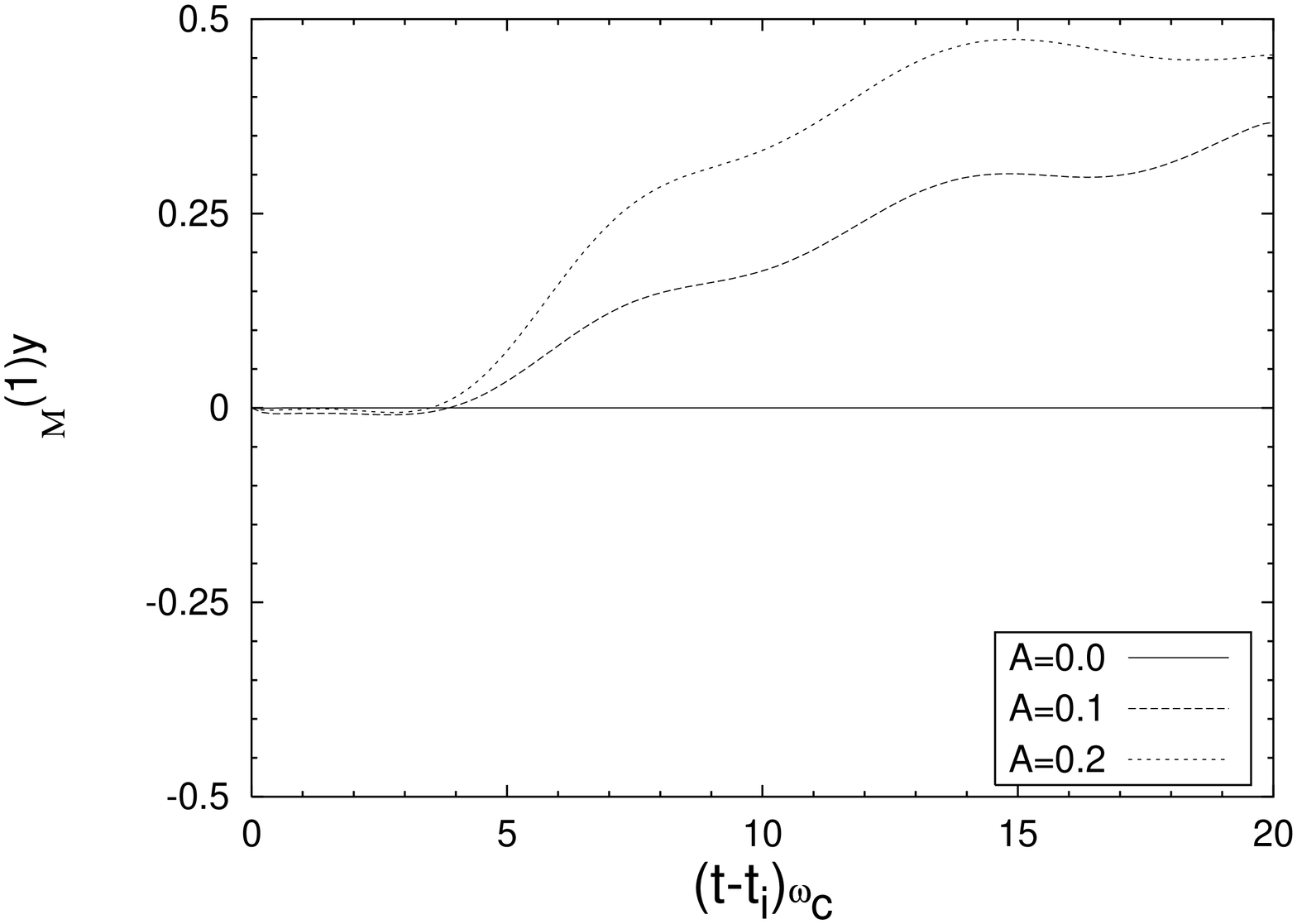,height=2.5in,width=2.5in,angle=270}
\epsfig{file=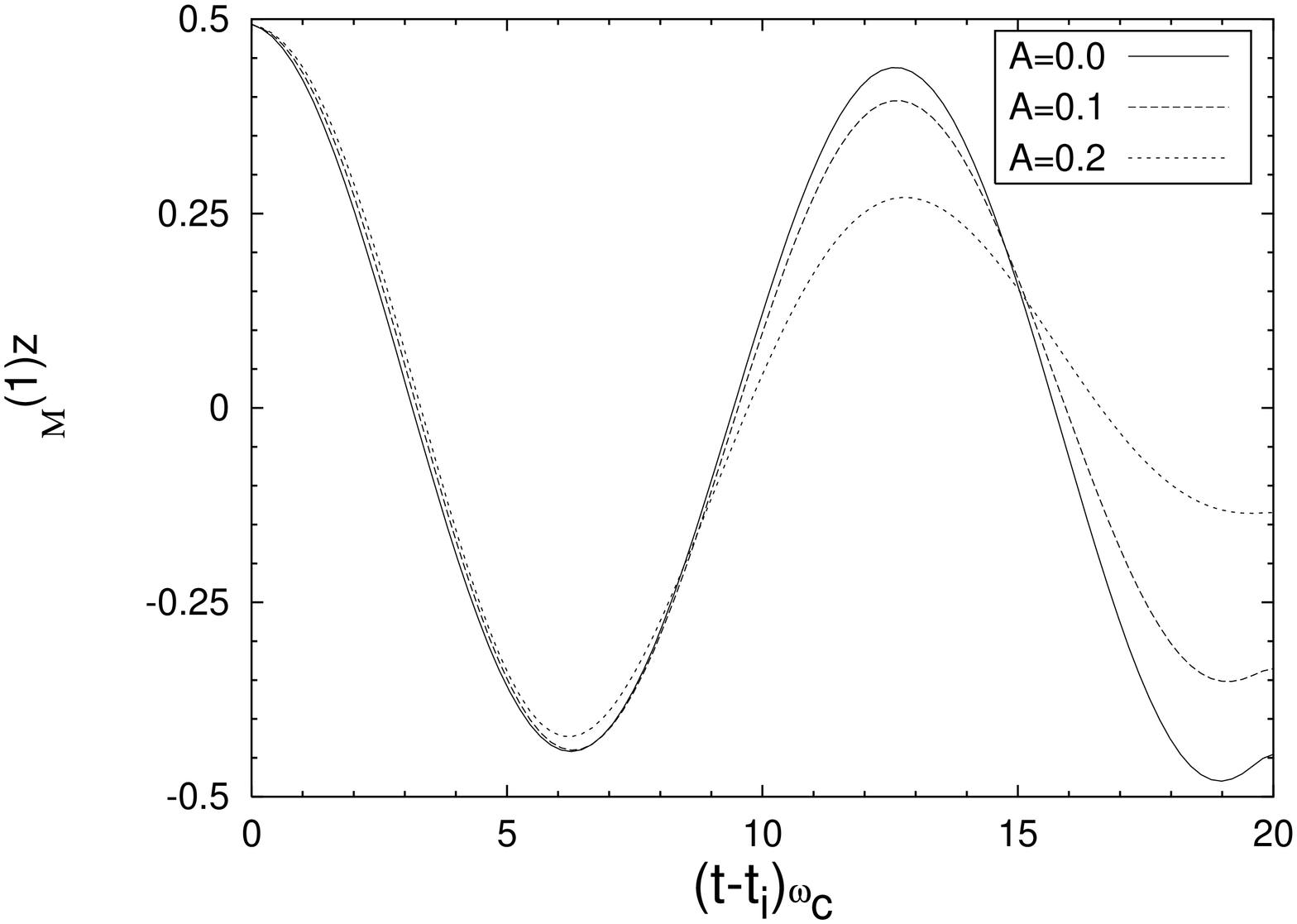,height=2.5in,width=2.5in,angle=270}
\end{center}
\caption{Time-dependence of one-spin magnetization after 
the $\pi /2$-rotation of the magnetic field
${\vec H}_{i}=(0,0,H)\rightarrow {\vec H}_{f}=(0,H,0)$
at different values of the spin-phonon coupling $A$.
Other parameters are fixed
at $s=0.5, H/\omega_c=0.5, T_i/\omega_c=T_f/\omega_c=0.1$.}
\label{foobar:fig4}
\end{figure}

\newpage

\section{ONE SPIN IN A TIME-DEPENDENT MAGNETIC FIELD}

As a first example we consider the problem of
the behavior of the magnetization of 
one spin in the presence of the
homogeneous time dependent magnetic field ${\vec H(t)}=(0,0,H_{z}(t))$.
We will consider the effect of performing spin rotations by $\pi$ or
$\pi/2$.
Let us consider first the case of magnetic field flipping.
It is supposed that the system is in equilibrium at initial
time $t=t_{i}$ at temperature $T_{i}$
and at $t=t_{i}+\delta$ the magnetic field
suddenly changes its direction: ${\vec H}\rightarrow -{\vec H}$.

Also at this time the coupling to the phonons at temperature $T_{f}$
is switched on,
otherwise the relaxation to equilibrium will not take place.
The heat bath is necessary to get magnetization relaxation to equilibrium
with the magnetic field $-{\vec H}$. Otherwise, the magnetization will precess
around the old field ${\vec H}$, as it follows from the equation
for magnetization in the case of no spin-phonon coupling:
$$
i\frac{d}{dt}{\vec M}=-{\vec M}\times{\vec H}.
$$

Let us get the equation for $\phi_{i}^{(0)}(t)$ when
there is spin-phonon coupling. 
Since the magnetic field is homogeneous in space, 
the field $\phi_{i}^{(0)}(t)$ 
and the magnetization ${\vec M}_{i}^{(0)}(t)$
do not depend on site $i$, and 
the mean-field equation (\ref{phi}) can be written as
\begin{equation}
{\vec \phi}_{i}^{(0)}(t)={\vec H}_{i}(t)+i
\int_{C_{1}+C_{2}}dt'\Delta (t,t'){\vec M}_{i}^{(0)}(t'),
\label{phi2}
\end{equation}
This last equation is equivalent to
\begin{equation}
{\vec \phi}_{i}^{(0)}(t)={\vec H}_{i}(t)-2
\int_{t_{i}}^{t}dt'\Im \alpha (t-t'){\vec M}_{i}^{(0)}(t'),
\label{phi3}
\end{equation}
where 
$$
\Im \alpha (t-t')=-A\int_{0}^{\infty}d\omega \omega^{s} e^{-\omega/\omega_{c}}
\sin (\omega (t-t'))
$$

This is a simple renormalization of the magnetic field induced by the heath 
bath and does not depend on the heat bath temperature $T_{f}$. 
Therefore, a higher correction beyond the saddle-point level must be 
considered. 

We shall study the behavior of the magnetization $M^{(1)}$
of one spin calculated by expressions (\ref{M1detail}) and (\ref{B}).
For one spin there are no site indices in these expressions
and the formula for the magnetization can be simplified. Let us limit
our consideration to the linear term in $\Delta$, which assumes that the heat
bath coupling is weak.
Then, the expression for the $z$-component
of the magnetization has the following form: 
\begin{eqnarray}
M^{z(1)} (t)\simeq M^{(0)}(t) \nonumber \\
+\frac{i}{4}\int_{C_{1}+C_{2}}dt_{1}dt_{2}
\Delta^{nn}(t_{1}-t_{2})
\varepsilon^{nr_1r_2}\varepsilon^{nn_1n_2}\varepsilon^{kk_1k_2}
\nonumber \\
\times 
[ G^{k_{2}n_{1}}(t,t_{1})G^{n_{2}r_{1}}(t_{1},t_{2})G^{r_{2}k_{1}}(t_{2},t)
\nonumber \\
+G^{k_{2}r_{1}}(t,t_{2})G^{r_{2}n_{1}}(t_{2},t_{1})G^{n_{2}k_{1}}(t_{2},t)
] .
\label{M1spin}
\end{eqnarray}
Since we are interested in the correction
to the magnetization linear in $\Delta$, the Green's function $G(t_{1},t_{2})$ 
in equation (\ref{M1spin})
can be calculated at ${\vec \phi}^{0}(t)=-{\vec H}(t)$.
When the magnetic field flipping is applied, the expressions for free Green's
functions (\ref{G>}) and (\ref{G<})
remain correct with the only substitution $H^{z}=-H$
in the complex exponents. 
The Fermi-factors 
in these expressions do not change with
 the magnetic field flipping, since they are defined by the initial conditions,
but the time exponentials, which define the dynamics of the spins,
are changed accordingly. The overall expression does not get modified because 
both fields are along the same direction.

Assuming, for simplicity, that only the $x$-component of the spin is coupled 
to the heat bath ($\Delta^{nn} \sim\delta^{nx}$),
substituting (\ref{G}), (\ref{G>}) and (\ref{G<}) in (\ref{M1spin}) and 
integrating over $t_{1}$ and $t_{2}$ leads to the following expression
for the $z$-component of the magnetization.
\begin{eqnarray}
M^{z(1)}(t)=M^{(0)}-A\int_{0}^{\infty}d\omega \omega^{s} 
e^{-\omega /\omega_{c}}
\nonumber \\
\times\{ [\frac{1}{2}+M^{(0)}\coth (\frac{\beta_{f}\omega}{2})]
  \frac{\sin^{2}\frac{(\omega +H^{z})(t-t_{i})}{2}}{(\omega +H^{z})^{2}}
\nonumber \\
+
   [-\frac{1}{2}+M^{(0)}\coth (\frac{\beta_{f}\omega}{2})]
  \frac{\sin^{2}\frac{(\omega -H^{z})(t-t_{i})}{2}}{(\omega -H^{z})^{2}}
\},
\label{Mspinsol}
\end{eqnarray}
where $M^{(0)}=\frac{1}{2}\tanh (\frac{\beta_{i}H}{2})$ is the initial 
magnetization.
Below we shall use an approximation $e^{-\omega /\omega_{c}}\simeq
\theta (\omega_{c}-\omega )$ in the numerical calculations, for convenience.
Obviously, expression (\ref{Mspinsol}) is correct for small times $(t-t_{i})$,
when the perturbation theory is valid, which is the regime where the magnetic 
field flipping is important. 
The time dependence of the magnetization for different types of the phonon
heat bath is presented in Fig.2.
As it follows from this figure the phonon density of states plays an important
role in the spin dynamics.
The behavior of the magnetization  after lowering the temperature 
for the cases of magnetic field flipping and constant magnetic field
is presented in Fig.3 for comparison.
The magnetization decreases as $(t-t_{i})^{2}$
at short times in both cases and is a linear function
of $t-t_{i}$, when $(t-t_{i})\gg 1/|H|$ (see (\ref{Mspinsol}) and
asymptotic expression (\ref{delta}) for long times).

Let us now consider a qualitative scheme of
how to obtain from expression (\ref{Mspinsol}) the relaxation 
of the magnetization to the equilibrium value at $T=T_{f}$.
In the case when $\Delta t\equiv (t-t_i)\gg 1/\omega_c , 1/H_{z}$ 
the following relation can be used
\begin{equation}
\frac{\sin^{2}((\omega\pm H^{z})(t-t_{i})/2)}{(\omega\pm H^{z})^{2}}
\simeq \frac{\pi}{4} (t-t_{i})\delta (\omega\pm H^{z}).
\label{delta}
\end{equation}
Then, equation (\ref{Mspinsol}) can be transformed to
\begin{eqnarray}
M^{z(1)}(t_i+\Delta t)=M^{z(0)}(t_i)
\nonumber \\
-\lambda \Delta t
(M^{z(0)}(t_i)-M_{f}^{z(0)}),
\label{int1}
\end{eqnarray}
where
\begin{equation}
\lambda =\frac{\pi A |H^{z}|^{s}}{8|M_{f}^{z(1)}|}e^{-\omega /\omega_{c}}
\label{lambda}
\end{equation}
and the equilibrium value of the magnetization at $T=T_f$ is
\begin{equation}
M_{f}^{z(1)}=\frac{1}{2}\tanh (\frac{\beta_{f}H^{z}}{2}).
\label{Mf}
\end{equation}

In the case when $\Delta t$ is not too large, but much bigger than
$1/\omega_c$ and $1/H$, and taking $t_i$ as 
arbitrary, equation (\ref{int1}) 
can be transformed into the differential equation
\begin{equation}
\frac{d}{dt}M^{z(1)}(t)=-\lambda 
(M^{z(1)}(t)-M_{f}^{z(1)})
\label{int2}
\end{equation}
with the initial boundary condition
\begin{equation}
M^{z(1)}(t_{i})=M^{z(0)}=\frac{1}{2}\tanh(\frac{\beta_{i}H}{2}).
\label{bc}
\end{equation}
The solution of (\ref{int2}), (\ref{bc}) is
\begin{equation}
M^{z(1)}(t)=M_{f}^{z(1)}+(M^{(0)z}-M_{f}^{z(1)})
e^{-\lambda (t-t_i) }.
\label{M1sollongt}
\end{equation}
As it follows from equation (\ref{M1sollongt}), the magnetization
approaches its equilibrium value exponentially,
and the relaxation time $1/\lambda$ 
is defined by the final temperature, magnetic field, 
and by the heat bath parameters.

As another example let us now consider the time-dependence of the magnetization
under the effect of a $\pi /2$-rotation of the magnetic field:
${\vec H}_{i}=(0,0,H)\rightarrow {\vec H}_{f}=(0,H,0)$. 
The dependence calculated by means of expression
(\ref{M1spin}) in the case of the lowest order spin $x$-component-phonon
coupling is presented in Fig.4.
The magnetization is precessing and its direction 
is changing towards the new direction of the magnetic field
in the case of non-zero spin-phonon coupling.
 
\section{THE CASE OF MANY SPINS}

In the case of many spins the time relaxation dynamics is complicated
by the spin-spin interaction. Let us simplify the problem by taking into 
account the spin-spin interaction at the mean-field level. 
The spin Hamiltonian in the case of the magnetic field in $z$-direction
can be written as
$$
H_{spin}(t)=-2dJM^{(0)}S^{z}(t)-H^{z}(t)S^{z}(t),
$$
where $d$ is dimensionality of the system and 
the initial magnetization $M^{(0)}$ is defined by the equation
\begin{equation}
M^{(0)}=\frac{1}{2}\tanh (\frac{\beta_{i}(H+2dJM^{(0)})}{2}).
\label{MMF}
\end{equation}
Therefore, the problem is reduced to one spin in the effective field
${\tilde H} =H^{z}+2dJM^{(0)}$. The formula (\ref{Mspinsol})
with the substitution $H^{z}\rightarrow H^{z}+2dJM^{(0)}$ describes
the time-dependence of the magnetization of the Heisenberg model after
the magnetic field flipping. 
The time-dependence of the magnetization for the spin system at 
different values of $dJ$ is presented in Fig.5.
As it follows from this Figure, the time-dependence of the magnetization
strongly depends on the relation $|H|/dJ$ keeping the other parameters fixed.
Let us analyse the expression (\ref{Mspinsol}) at quite
long times, when the relation (\ref{delta})
can be applied. Either the first or the second terms in the curled brackets
contribute to the integral in (\ref{Mspinsol}), when the sign
of ${\tilde H}=-|H|+2dJM^{(0)}$ is negative or positive, respectively.
\begin{figure}% fig 5
\begin{center}
\epsfig{file=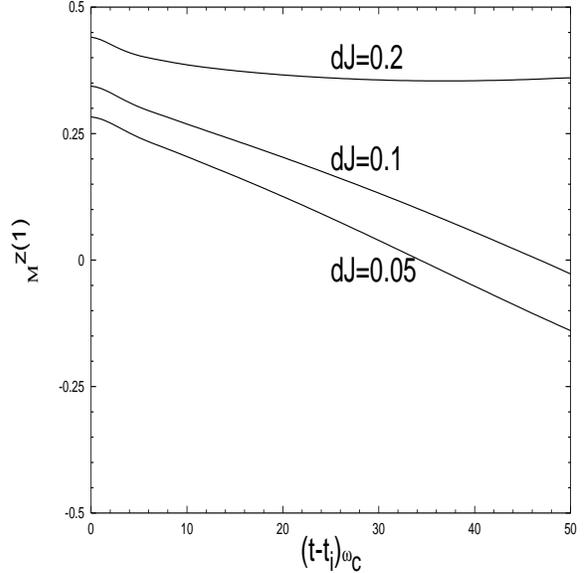,height=3.0in,width=3.0in,angle=270}
\end{center}
\caption{Time-dependence of the magnetization for spin system after magnetic
field flipping at different values
of $dJ$ (in units of $\omega_{c}$) and $A=0.025, s=0.5, 
H/\omega_c=0.1, T_i/\omega_c=0.1, T_f/\omega_c=0.02$.}
\label{foobar:fig5}
\end{figure}
This leads to different, negative or positive, corrections to the magnetization
due to the magnetic field flipping. Therefore, the equation ${\tilde H}=0$
defines the critical magnetic field value $H_{cr}$, below which the magnetic
field flipping does not lead to a change of the magnetization sign.   
Equation ${\tilde H}=0$ together with (\ref{MMF}) gives the expression, which 
connects $H_{cr}$ with $J$ and $T_{i}$:
$$
\frac{H_{cr}}{dJ}=\tanh (\frac{H_{cr}}{T_{i}}).
$$
The initial temperature dependence of $H_{cr}$ is presented in Fig.6.
As it follows from this Figure, at low temperatures, the critical value of magnetic field is $dJ$, i.e. it is of order of the ferromagnetic coupling energy,
which should be overcome to change the sign of the magnetization
of the system. 
The time-dependence of the magnetization for the spin system 
at different values of $H$ is presented in Fig.7.
As it follows from this Figure, the magnetization does not flip
when the magnetic field is lower than the critical value 
$H_{cr}\simeq 0.9575dJ= 0.1915\omega_{c}$ at given temperature
$T_{i}=0.1\omega_{c}$ (see Fig.6). In this case the magnetization
oscillates with a reasonably large amplitude, similarly
to the case of the $\phi^{4}$-theory \cite{Cao1,Cao2} and does not
relax to the overturned magnetization. In the $\phi^{4}$-theory a vector
self-interacting theory is considered. In our case the interaction part is
due to the spin interactions.

\begin{figure}% fig 6
\begin{center}
\epsfig{file=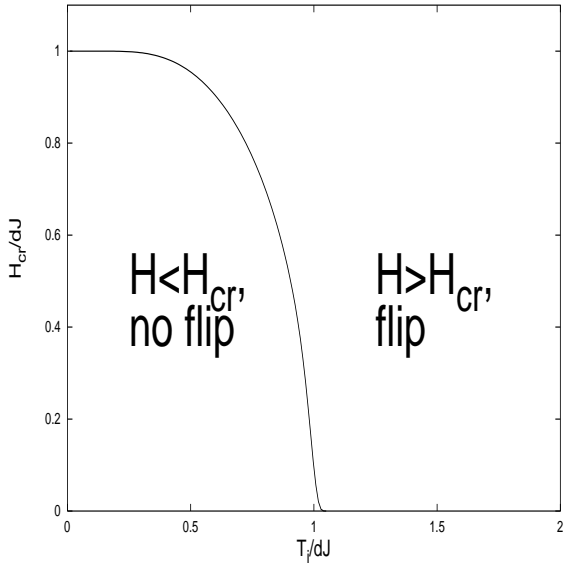,height=3.0in,width=3.0in,angle=270}
\end{center}
\caption{Temperature-dependence of $H_{cr}$. Mean-field critical
temperature of ferromagnetic transition is $T_{c}=0.5dJ$.}
\label{foobar:fig6}
\end{figure}

Finally, in Fig.8 we present the normalized magnetization deviation
$(M^{(0)}-M(t))/[(t-t_i)\omega_{c}]$. The rate of change of the magnetization
 corresponds to the  Fermi golden rule
for the scattering cross section. After an initial rapid increase the rate
converges to a constant valid for intermediate times, as well
known. For very large times the perturbation theory is no longer valid. In this
regime we have to consider higher order corrections which complicates the
solution. This will be considered in a future publication.  

\begin{figure}% fig 7
\begin{center}
\epsfig{file=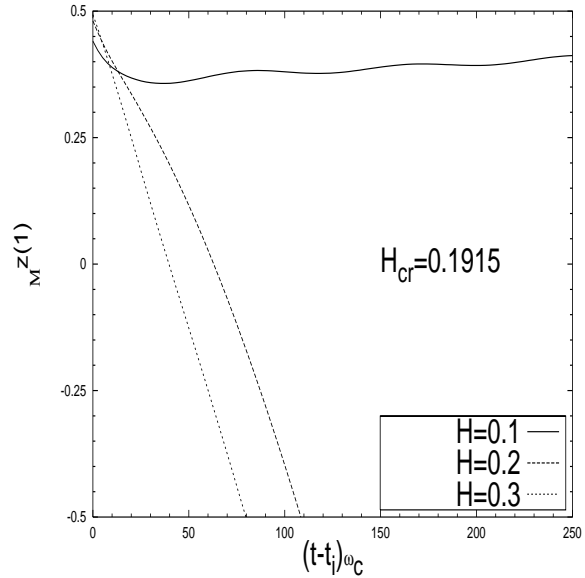,height=3.0in,width=3.0in,angle=270}
\end{center}
\caption{Time-dependence of magnetization for spin system at different values
of $H$ (in units of $\omega_{c}$) and $A=0.025, dJ/\omega_c=0.2, 
s=0.5, T_i/\omega_c=0.1, T_f/\omega_c=0.02$. Critical value
of the magnetic field in this case is $H_{cr}\simeq 0.1915\omega_{c}$.}
\label{foobar:fig7}
\end{figure}
 
\section{CONCLUSION}

The problem of the time-dependence of the magnetization of a spin
system in a time-dependent magnetic field is interesting both
from the point of view of its practical applications, and from the theoretical
point of view. In this paper we have studied the behavior of the Heisenberg
ferromagnet coupled to a heat bath in a time dependent external magnetic field.
In particular, we considered the case when
its direction is suddenly changed. As expected the behavior of the system strongly depends on the phonon bath properties. 

It has also been shown that there is a critical value of the magnetic 
field below which the magnetization
does not relax to the equilibrium value after the magnetic field flipping.
This situation is analogous to the $\phi^{4}$-theory case, where
also exists a critical value of the external source \cite{Cao1,Cao2}.
\begin{figure}% fig 8
\begin{center}
\epsfig{file=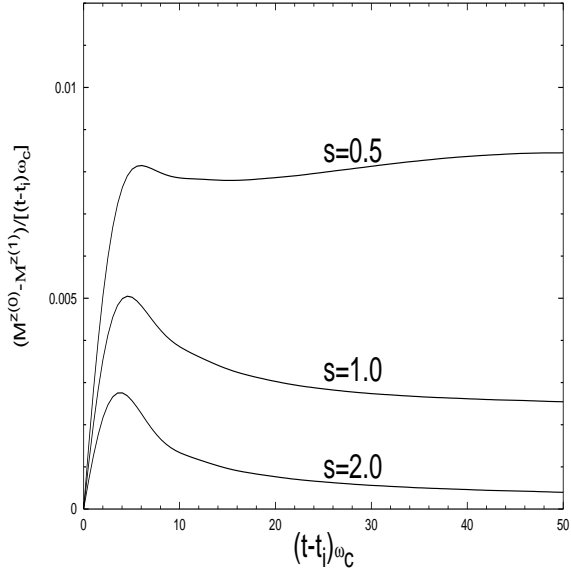,height=3.0in,width=3.0in,angle=270}
\end{center}
\caption{Time-dependence of the rate of change of the magnetization for the spin system 
after magnetic field flipping 
at different values of $s$ and $A=0.025, dJ/\omega_c=0.05, 
H/\omega_{c}=0.1, T_i/\omega_c=0.1, T_f/\omega_c=0.02$.}
\label{foobar:fig8}
\end{figure}
In this work we considered a microscopic description for the vector
spin system directly instead of using an effective bosonic theory from the
start. Representing the spin operators by Majorana fermions we constructed
the path integral representation for the spin system. Also, we used a physical
way to introduce the relaxation to equilibrium coupling the system to a heat
bath. For simplicity we used a phonon heat bath. The spin-phonon coupling
can be integrated exactly introducing an effective time-dependent interaction
between the spins. Afterwards we bosonized the model via a Hubbard-Stratonovich
transformation leading to an effective bosonic theory (after integrating out
the Majorana fermions) which has some resemblances with the $\phi^{4}$-theory
but with significant differences like time-dependent coefficients and the
introduction of cubic terms (like $\phi^{3}$) due to the presence of a three
field vertex which originates in the three-spin correlation function which does
not vanish in general due to the nature of the spin commutation relations. 

At the same time there are several questions which remain open like 
the interpolation between the short time and the long time behaviors. 
Also, in the problem of quantum spinodal decomposition, the possibility
of the formation of a magnetic bubble of the stable phase immersed 
in a background of the unstable phase and the determination of its radius 
time-dependence is an interesting open problem.
These and some other questions are planned to be studied in the nearest
future where a detailed comparison of the Heisenberg model with the 
$\phi^{4}$-theory will be carried out.

\end{document}